\documentclass[useAMS,usenatbib,a4paper]{mn2e}
\usepackage{aas_macros}
\usepackage{graphicx,amssymb}
\usepackage[usenames]{color}

\title{Photometric Selection of $z\sim5$ Lyman Break Galaxies in the ESO Remote Galaxy Survey}

\author[L. S.~Douglas et~al.]{L.S.~Douglas$^{1,2}$, M.N. Bremer$^1$, E.R. Stanway$^1$  M.D. Lehnert$^2$ \& D. Clowe$^3$ \\
$^1$H H Wills Physics Laboratory, Tyndall Avenue, Bristol, BS8 1TL, UK\\
$^2$Laboratoire d'Etudes des Galaxies, Etoiles, Physique et Instrumentation GEPI, Observatoire de Paris, Meudon, 92195, France\\
$^3$Department of Physics and Astronomy, Ohio University, 251B Clippinger Lab, Athens, OH 45701, USA}

\begin{document}

\date{Accepted . Received ; in original form }

\pagerange{\pageref{firstpage}--\pageref{lastpage}} \pubyear{}

\maketitle

\label{firstpage}

\begin{abstract}

We describe the selection of a sample of photometrically-defined
Lyman break galaxies (LBGs) at $z\sim5$ using the multi-wavelength imaging data of
the ESO Remote Galaxy Survey (ERGS). The data is drawn from ten
widely-separated fields covering a total sky area of 275
arcmin$^2$. Starting with a simple colour ($R-I>1.3$) and magnitude
($I<26.3$) cut to isolate the Lyman break and then refining the sample
by applying further optical and near-infrared photometric criteria we
identify a sample of 253 LBG candidates. We carefully model the
completeness of this sample and the factors that affect its
reliability. There is considerable overlap between this sample and a
spectroscopically-confirmed sample drawn from the same survey and this
allows us to determine the reliability of the optical photometric selection
($\sim 60$ per cent) and to show that the reliability can be
significantly improved (to $\sim 80$ per cent) by applying near-infrared
waveband criteria to exclude very red contaminants. Even
this high level of reliability may compromise some statistical studies
of LBG properties. We show that over 30 per cent of the highest
reliability candidates have multiple UV-luminous components and/or disturbed morphology
in HST imaging, though it
is unclear whether this represents multiple interacting/merging
sources or individual large sources with multiple UV bright
regions. Using this sample we confirm that the
normalisation of the bright end of the $z=5$ UV luminosity function
(down to $M^*$) is lower than the same at $z=4$ by a factor of
3. Using a Schechter fit we determine $M^*_{UV}=-20.9\pm{0.2}$. We discuss
whether it is reasonable to expect the UV luminosity function to
follow a Schechter function, given the UV emission is short-lived and
stochastic, and does not necessarily trace the underlying mass of the
galaxy.

\end{abstract}


\section{Introduction}
\label{sec:intro}

Until the turn of the century, very few spectroscopically-confirmed
galaxies at $z\sim 5$ and above had been identified and fewer still
studied in any detail. Given that this redshift range covers the first
billion years after the Big Bang, this meant that there was little
concrete evidence of how early galaxy formation proceeds. With the
advent of deep imaging with 8m ground-based telescopes, and with the
GOODS and HUDF projects carried out with HST, it has become possible to
photometrically identify distant galaxy candidates and potentially
confirm them through spectroscopy.

By extending the Lyman break technique \citep{guhath90,steidel92,steidel93},
to $z\sim5$ using $R,I$ and
$z-$band imaging, \cite{lehnert03} showed that with the right observational
setup it is now possible to securely identify multiple high redshift
galaxies in individual 8m telescope pointings. These Lyman Break
Galaxies (LBGs) are identified {\it via} their redshifted UV continuum
emission arising from strong unobscured star formation within the
systems. Subsequently, similar studies have been carried out by other
authors in order to select galaxies with redshifts above
$z=4.5$ \citep{ouchi04} and potentially to $z>6$ \citep[e.g.][]{
stanway03,stanway04,stanway08a}.

While some of these studies have included spectroscopic confirmation
of samples derived from photometry, many others have not.  These have
relied on the high success rate of spectroscopic confirmation in
previous work \citep[ e.g.][confirmed six of a sample of twelve
  candidates in a 40 arcmin$^2$ field to be at $4.8<z<5.8$]{lehnert03}
to argue that photometric selection was reliable enough to carry out
statistical studies on photometric samples. However, as shown by
\cite{stanway08b}, the reliability and completeness of such samples
depends on a complex combination of factors, particularly the filter
set and CCD response used to carry out the imaging. In the absence of
spectroscopy, other data is required to help exclude contaminating
galactic stars and $z\sim 1$ elliptical galaxies from photometric
samples, such as HST and/or deep near-IR ground-based imaging. One of
the main advantages of purely photometric samples is that they can
cover a relatively large area of sky in comparison to spectroscopic
samples. However, given the difficulty of obtaining HST and near-IR
data of sufficient depth over a large sky area, the use of photometric
samples are often compromised.

For this reason, despite the apparent ease of defining a photometric
sample, there are distinct advantages in using a sample which has
clear spectroscopic evidence of reliability when attempting to
understand the statistics and other properties of star forming
galaxies at $z>5$. Accurate determinations of their clustering statistics,
luminosity functions, star formation histories, stellar population
properties and contribution to reionization all require an
understanding of the contaminants in the sample being used for the
study. Adding to this the need for accurate redshifts in order to carry
out many detailed studies of individual high redshift systems, the
need for a well-defined photometric sample of $z\sim 5$ LBGs with good
spectroscopic follow-up is clear.

\cite{lehnert03} showed that it was possible to obtain ground-based
samples of such objects with a reasonable level of spectroscopic
confirmation. However, the study covered a comparatively small area
of sky and so could only be used to determine the simplest of
statistical properties for such sources. In more recent years larger areas
have been covered by spectroscopic programs e.g. \citet{ando07} and 
\citet{vanzella09} which confirmed 10 and 32 high redshift galaxies 
respectively. Although these surveys are important to the understanding
of the selection of high redshift galaxies and their properties, they often 
only target the brightest candidates spectroscopically or do not uniformly
probe the corresponding photometric sample. One of the goals
of this survey, described here and in future papers, is 
to observe many high redshift LBG candidates over a range of luminosities 
and colours to discover if the relatively high completeness seen in previous studies
is applicable at fainter magnitudes. Although the level of contamination of low redshift
Lyman break galaxy samples are often very low, {\it{e.g.}} \citet{steidel03}, such a
low level cannot be assumed for higher redshift samples. The differences
in the selections functions and colours used mean that the two samples are  
contaminated by completely different galactic and extragalactic populations. 

To address this, and other, goals we have carried out a large programme 
of imaging and spectroscopic observations which
cover a sky area fourteen times larger than the original \citet{lehnert03} study
including an expanded region of $\sim 160$ arcmin$^2$ centered on the
original \cite{lehnert03} pointing. This paper introduces the imaging data sets used to
generate a sample of $z\sim5$ LBG candidates which were followed-up by
the same spectroscopic setup used in the original \cite{lehnert03}
work. Also it describes a photometric sample of $z\sim5$ LBG
candidates with well-determined reliability and completeness which can
be used for statistical study, such as a determination and discussion
of the $z\sim 5$ LBG UV luminosity function presented here and a brief comparison
of the results of a single $R-I$ colour cut to that of a sample selection with
an additional $I-z$ criterion. 
 
The data presented here are drawn from from ten $\sim40$ arcmin$^2$
fields widely-separated on the sky. Observations were carried out as
part of an ESO Large Programme, ``The ESO Remote Galaxy Survey'', or
ERGS (PI M. Bremer, ID 175.A-0706). This capitalised on existing deep
optical imaging of multiple fields already obtained by the ESO Distant Cluster Survey (EDisCS)
project \citep{white05}, itself an ESO Large Programme (PI S. White,
ID: 166.A-0162). The resulting spectroscopic sample of ERGS
contains many tens of high redshift galaxies with secure redshifts \citep{douglas09}.

The structure of this paper is as follows. In section
\ref{sec:observations} we discuss the imaging data that was used as a
basis for this work and briefly describe follow-up spectroscopic observations. In section \ref{sec:select} and \ref{sec:select} we discuss the
photometry and how the candidates were selected. We then discuss the final sample in section \ref{sec:highz} 
and the estimated contamination fraction. The
completeness of this sample is described in section \ref{sec:completeness}. The morphology and observed UV luminosity
function for the high redshift LBG candidates is considered in sections \ref{sec:morph} and \ref{sec:nc}

We adopt the standard $\Lambda$CDM cosmology, i.e. a flat universe
with $\Omega_{\Lambda}=0.7$, $\Omega_{M}=0.3$ and $H_{0}=70
{\rm km\,s}^{-1}\,{\rm Mpc}^{-1}$.  All photometry was determined in
the AB magnitude system \citep{oke83}.

\section{Observations and Data}
\label{sec:observations}

In order to select a sample of objects containing $z\sim5$ LBGs all
that is needed in principle is sufficiently-deep $R$ and $I-$band
imaging. Selecting a flux limited sample in $I$ containing objects
with an $R-I$ colour redder than a particular value (usually $R-I>1.3$
or $R-I>1.5$) should identify a sample of $z\sim5$ LBGs with a
completeness that depends on the statistical quality of the
data. However, to {\it cleanly} select a sample of high redshift
galaxy candidates without being swamped by contaminating cool Galactic
stars and lower redshift galaxies, a wide range of multi-band deep
imaging is required.

\subsection{Imaging}

The data set used in this study includes optical
imaging using the $V$, $R$, $I$ and $z$ bands, near infrared imaging
in the $J$ and $K_s$ bands and complementary high resolution $I$-band
imaging from the Hubble Space Telescope (HST) of ten widely separated fields each of size 
$\sim 40$ arcmin$^2$. The high Galactic latitudes of these fields
result in minimal Galactic extinction effects and the original
selection of the fields to observe $z\sim 0.7$ clusters produced an
effectively random sampling of the sky at $z\sim5$, reducing the effects
of cosmic variance.

The majority of the imaging data were taken as part of EDisCS
\citep[]{white05}, a photometric and spectroscopic survey of galaxy
clusters selected from the Las Campanas Distant Clusters Survey
\citep{gonzalez01}. The observing programme involved deep optical
imaging using FORS2 at the VLT, near-infrared imaging using SOFI at
the NTT, and high resolution imaging using the ACS camera on HST. The
fields were observed with FORS2 in the $V$, $R$, and $I$ bands for
approximately 2 hours each, with a field of view after dithering of
$6.5'\times6.5'$. In the near-infrared, the fields were observed using
SOFI on the NTT in the $J$- and $K_s$-bands for at least 300 and 360
minutes respectively. The effective field of view of the near IR
observations was typically $5.4'\times4.2'$, consequently only around
50 per cent of each optical field was covered. This ground based
imaging was complemented by 80 orbits of HST/ACS imaging
\citep{desai07}. Nine of the ten fields were observed for one orbit
over the whole field and an additional four orbits in the central
region using the F814W filter, with over 80 per cent of each
ground-based optical field being covered by HST data. With typical
seeing conditions for the ground based observations between 0.5$''$
and 0.8$''$, the 2$\sigma$ depth in a 2$''$ diameter aperture of each
image was typically 28.1 in the $V$-band, 27.9 in the $R$-band, 27.1
in the $I$-band, 24.5 in the $J$-band and 23.7 in the
$K_s$-band. Details can be found in Table \ref{depth_tab}. The depths of the
images were calculated by measuring the noise properties of multiple randomly placed 
2$''$ apertures in background areas of the images.

These images were reduced using standard techniques for bias and bad
pixel removal and flat fielding. The software IMCAT $\it findpeaks$
was used to detect local minima in images smoothed with a 1$''$
Gaussian to create a catalogue of local sky measurements. These
measurements were then fitted to a bi-cubic polynomial and subtracted
from the unsmoothed image. The images were also aligned with known
stars in the USNO catalogue to correct for camera distortions and the
linear offset between the fields, preserving the surface
brightness. Specific details of the data reduction producing the
calibrated images used in this study can be found in \citet{white05}.

\begin{table*}
\caption{Magnitude limits for 2$\sigma$ detections in a 2$''$ circular aperture. The mean weak lensing corrections for each field, as calculated from the maps of \citet{clowe06}, are shown in the last column. \label{tab:depth}}
\scriptsize
\label{depth_tab}
\begin{center}
  \begin{tabular}{|c|c|c|c|c|c|c|c|c|c|c|c|c|c|}
\hline
\bf Field& $\mathbf{V}$ & seeing & $\mathbf{R}$ & seeing & $\mathbf{I}$ & seeing & $\mathbf{z}$ & seeing & $\mathbf{J}$ & seeing & $\mathbf{K_{s}}$ & seeing & Lensing mag\\ \hline
J1037.9-1243 & 28.11 & 0.55$''$ & 28.04 &0.54$''$ & 27.00 &0.56$''$ & 26.17 &0.57$''$ & 24.61 &0.82$''$ & 23.93 &0.82$''$ & 0.12 \\ \hline
J1040.7-1155 & 28.08 & 0.65$''$ & 27.95 &0.72$''$ & 26.94 &0.62$''$ & 26.12 &0.77$''$ & 24.56 &0.70$''$ & 23.49 &0.68$''$ & 0.08 \\ \hline
J1054.4-1146 & 28.04 & 0.67$''$ & 28.07 &0.78$''$ & 27.04 &0.72$''$ & 26.03 &0.58$''$ & 24.32 &0.83$''$ & 23.53 &0.68$''$ & 0.24 \\ \hline
J1054.7-1245 & 28.07 & 0.79$''$ & 27.94 &0.77$''$ & 27.24 &0.50$''$ & 26.10 &0.62$''$ & 24.45 &0.82$''$ & 23.72 &0.78$''$ & 0.26 \\ \hline
J1103.7-1245 & 28.05 & 0.83$''$ & 28.01 &0.75$''$ & 27.19 &0.64$''$ & 26.23 &0.55$''$ & 24.38 &0.77$''$ & 23.56 &0.63$''$ & 0.20 \\ \hline
J1122.9-1136 & 28.12 & 0.71$''$ & 28.00 &0.70$''$ & 27.19 &0.65$''$ & 26.20 &0.55$''$ & 24.51 &0.79$''$ & 23.59 &0.61$''$ & 0.11 \\ \hline
J1138.2-1133 & 28.07 & 0.58$''$ & 27.86 &0.68$''$ & 27.14 &0.60$''$ & 26.14 &0.65$''$ & 24.59 &0.82$''$ & 23.69 &0.82$''$ & 0.17 \\ \hline
J1216.8-1201 & 28.05 & 0.68$''$ & 27.92 &0.72$''$ & 27.16 &0.60$''$ & 26.23 &0.50$''$ & 24.50 &0.76$''$ & 23.58 &0.67$''$ & 0.37 \\ \hline
J1227.9-1138 & 28.09 & 0.73$''$ & 27.89 &0.83$''$ & 27.12 &0.74$''$ & 26.27 &0.65$''$ & 24.62 &0.96$''$ & 23.67 &0.70$''$ & 0.08 \\ \hline
J1354.2-1230 & 27.92 & 0.70$''$ & 27.96 &0.70$''$ & 27.16 &0.66$''$ & 26.35 &0.45$''$ & 24.51 &0.90$''$ & 23.98 &0.74$''$ & 0.13 \\ \hline

\end{tabular}
\end{center}
\end{table*}

\subsubsection{Additional $z$-band Imaging}
\label{sec:zband}

The high redshift galaxies selected in this project are observed due
to their bright rest-frame UV emission, redshifted to $\sim 740$nm at
$z\sim5$, arising from vigorous star formation within the galaxies. As
young star forming galaxies, they are expected to have approximately flat
continuum emission ($f_\nu=$constant, an AB colour of zero) longward
of Lyman $\alpha$ which can be probed in the observed $z$-band. In the
absence of other effects, the $I$-$z$ colour can also be an indication of a
galaxy's redshift as the Lyman break moves through the $I$-band, ($5.0
< z < 5.8$). A high $I$-$z$ colour would indicate very little flux in
the $I$-band caused by the Lyman break located near the red edge of
the $I$-band filter. Thus, the addition of $z$-band imaging can improve
photometric selection of high redshift galaxy candidates and the ability to study
a candidate's properties.

Each of the ten separate fields were observed for two hours in the
$z$-band with FORS2 \citep{appen98} using the z\_Gunn filter and 2$\times$2 pixel binning. The observations were carried out in March and
December of 2005 and February and March of 2007. For each field,
forty-two exposures of two minutes each were taken at slightly
different dither positions, similar to the technique used in the near
infrared. Many short exposures are needed to minimise the increasing
night sky emission at redder wavelengths. The high background is
caused by line emission from the excited levels of the hydroxyl radical
OH$^{-}$. This airglow can vary on timescales as short as five minutes
so having many short exposures is beneficial. The fields have a mean
depth of 26.18 AB magnitudes ($2\sigma$ limit in a 2$''$ aperture) for an
84 minute observation. The observations were taken with the same
instrument as the other optical imaging and the exposures were planned
to match the field of view of the existing images. The reduction
of the data followed standard techniques using $\it{IRAF}$
software. Bias subtraction, flat-fielding, bad-pixel removal and
photometric calibration using standard star observations were all
performed in the standard manner.

Although the data were obtained with the same camera as that used to
obtain the $V,R$ and $I-$band datasets, in the time between the two
sets of observations, the CCD chips were upgraded from SiTE SI-424A
backside thinned devices to MIT/LL devices. This upgrade improved the
quantum efficiency of the instrument in the $z$-band, and almost
completely removed any fringing longward of $8000$\AA. It also changed
the pixel scale of the data (from 0.2 to 0.125 arcsec per unbinned
pixel), a change that influences how we carry out the photometry
discussed section \ref{sec:photom}.

\subsection{Spectroscopy}

In addition to the $z$-band imaging, the ERGS program spectroscopically 
observed a range of high redshift candidates and intermediate and low
redshift contaminating objects. Although the spectroscopic follow-up will
be presented in detail in a forthcoming paper, a brief summary is given as 
some of the results will be referred to in sections \ref{sec:select} and \ref{sec:highz}.

The high redshift galaxy candidates were followed-up spectroscopically
in a 100 hours VLT/FORS2 program described in \citet{douglas09}.
The 10 survey fields were observed using 20 spectroscopic masks on
VLT/FORS2 each containing 30 to 40 slitlets with a width of 1 arcsec and typical length of
10 arcsec. Observations were carried out in service mode between 
December 2005 and May 2007 during dark time with seeing conditions better
than 1 arcsec and clear or photometric conditions at an airmass less
than two. Each mask was observed for 3.6 hours divided into exposures
of 650 seconds nodded along the slit  to avoid the effect of pattern noise or bad
pixels and facilitate the best possible sky subtraction.  FORS2 was
used in spectroscopic mode using the R300I grating with the OG525
blocking filter to restrict the wavelength range to between
650-1000nm. The resulting pixel scale was 3.2\AA\ in the spectral
direction and 0.2 arcsec spatially and the spectral resolution was
$\sim$13\AA, measured from the width of the sky lines.

The data were reduced following the same
standard procedures as those described in \citet{lehnert03} and hence 
consistent with the data in the previous work. More details
of the spectroscopic data reduction and analysis will be presented in \citet{douglas09}.

\section{Photometry}
\label{sec:photom}

Object catalogues were initially created from the fully reduced and
calibrated images using the galaxy photometry package
SExtractor version 2.4.4 \citep{bertin96}. Magnitudes were determined
using both 2$''$ circular apertures and SExtractor defined MAG\_AUTO apertures, which gave
the best estimate of total magnitudes for individual objects. Aperture
corrections were determined for the circular apertures from
the photometry of relatively isolated stars (identified as such using
the HST imaging) using larger apertures. A comparison of the corrected
circular aperture and MAG\_AUTO aperture photometry for a range of
unresolved sources showed agreement within statistical uncertainty.

\begin{figure}
\begin{center}
\includegraphics[width=8cm]{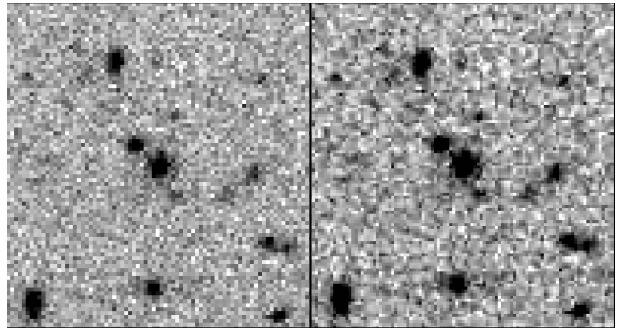}
\caption{The image on the left shows a portion of the fully reduced
  $z$-band image where no rebinning has taken place. The image on the
  right is of the same portion of the reduced $z$-band image rebinned
  to match the $I$-band image. The correlated pixel noise can be seen
  as a rectilinear pattern covering the background.}
\label{fig:zband}
\end{center}
\end{figure}

All high redshift candidates were expected to be
unresolved in the ground-based imaging \citep[they are found to have
half-light radii of around 0.1$''$-0.3$''$,][compared to typical
seeing of 0.7$''$]{bremer04}. The original intention was to apply the
MAG\_AUTO apertures to the other imaging bands with small
corrections (usually a few hundredths of a magnitude) for differences
in seeing in order to determine colours, or limits on colours. This
required resampling the $z, J$ and $K_{s}$-band images to the same pixel-scale and
astrometry as that of the $V, R$ and $I$-bands. Unfortunately the resampling
significantly correlated the noise in the
background, particularly in the $z$-band. This had a significant effect on the photometry
of the faint ($z>25$) LBG candidates (see figure
\ref{fig:zband}). Consequently, all subsequent photometry was carried
out using 2$''$ circular apertures on images which have not been
resampled, taking into account the appropriate aperture
corrections. This resampling did not affect the $V$, $R$ and $I$
images so we were able to additionally select objects which had
colours that fell into our selection criteria in the MAG\_AUTO photometry for these
bands. While not part of our main photometric sample discussed here,
these were included on spectroscopic slit masks where there was
available space.

SExtractor was used in two image mode, assigning the $I$-band image as
the detection image with which to define the photometric apertures for
the three bluest images, $V$, $R$ and $I$-band as these images were taken with the same pixel scale and central pointing. 
Objects were identified by a flux excess of
more than 1.5$\sigma$ over at least four consecutive pixels. For the redder bands the $I$-band image was
used as a detection image to define the centers of the
apertures. These RA and Dec positions were used to place appropriately
scaled 2$''$ apertures on the $z$, $J$ and $K_{s}$-band images.

SExtractor does not carry out photometry of an arbitrary list of
celestial coordinates. Consequently, this latter photometry was
carried out using a modified version of the {\rm APER} routine in
the IDL astronomical library. In each case the photometry was carried
out on images which had been background subtracted by SExtractor, 
using the {\rm APER} routine without any background subtraction, in
order to mimic as closely as possible the photometry carried out in
the bluer bands. The code was checked by comparing the resulting photometry of
the $I$-band frame with that from SExtractor, both giving consistent
results.

The resulting catalogue containing $V$-, $R$-, $I$-, $z$-band photometry 
for all objects detected in the $I$-band, along with $J$ and $K_{s}$ photometry 
for those objects covered by the near-IR imaging, was
then corrected for Galactic extinction \citep{schlegel98} producing the final photometry
catalogue used for high redshift candidate selection. The
uncertainties in the measurements were assessed by placing artificial
galaxies of known size and flux randomly on the science image and then
recovering them using the above methods. It was found that the measured
errors approached the theoretical Poissonian statistical error. Had an
optimised aperture been used, such as the MAG\_AUTO isophotal apertures used by
SExtractor, the measured errors would have reproduced the statistical
errors. In cases where objects were undetected
in a single band, the 2$\sigma$ magnitude limit in a 2$''$ diameter
aperture was used to determine colours.

These fields were originally chosen to study $z\sim0.8$ galaxy groups
and clusters as part of the EDisCS project \citep{white05}. Each field
contains a clustered system at $0.6<z<0.8$ with velocity dispersions
generally between $300<\sigma<900$kms$^{-1}$. One has a velocity
dispersion of $\sim 1000$kms$^{-1}$. These systems have a weak
lensing effect across the fields and potentially a strong lensing
effect over a small area. We used the lensing models derived by
\citet{clowe06} to determine the position-dependent
amplification for sources at $z\sim 5$ in order to determine the
magnitudes of the sources in our (observed) flux limited catalogue in
the absence of the intervening systems. These corrections were
typically of $\sim 0.1$ magnitudes and we found no strongly-lensed
 $z\sim 5$ candidates. The lensing effect was achromatic, changing the 
total magnitudes of objects but not their colours.

\section{LGB Candidate Selection}
\label{sec:select}

To take advantage of the near-IR imaging within this data set we use the following
selection method, described step-by-step, to produce a robust sample of LBG candidates. The optical 
$V$-, $R$- and $I$-bands are used to select a flux limited sample with a continuum
break which, if identified as the Lyman break, places the sources at $z>4.8$ (described in 
sub-section \ref{sec:optical}). The
near-IR imaging then becomes critical in identifying common low redshift stars and
galaxies with colours red enough to satisfy the optical criteria. Having described
such interlopers in sub-section \ref{sec:interlopers}, sub-section \ref{sec:nearir} discusses the impact
of the additional near-IR criteria. This leads to the final high redshift LBG candidate
sample in section \ref{sec:highz} which includes a discussion on the contamination fraction
estimate from spectroscopic observations of a subset of the sample.

\subsection{Optical Selection Criteria}
\label{sec:optical}

In order to identify objects with the characteristic spectral break
between the $I$- and $R$-band of high redshift star forming galaxies, the selection criteria of $I<26.3$ and
$R-I>1.3$ were imposed to create a catalogue of objects with a
significantly large spectral break or a non-detection in the
$R$-band. An additional condition of $V>27.5$ was added as this band
samples light around the Lyman limit at $z\sim5$.  As such, very
little, if any, flux is expected at these wavelengths. Although
applying this $V$-band cut will exclude a few per cent of the high redshift galaxies
(typically those at redshifts $4.7<z<5.0$ and magnitudes $I<25.5$), it has
the effect of decreasing the sample size by about 40 per cent and so
substantially decreases the contamination fraction (see simulations in section
\ref{sec:completeness}).

Including only those areas which are covered by all imaging bands ($V$-band to $K$-band) and
excluding regions near bright sources, a total of 319 objects were identified over the 10
survey fields covering an area of 275 arcmin$^2$.

\begin{figure}
\begin{center}
\includegraphics[width=8cm]{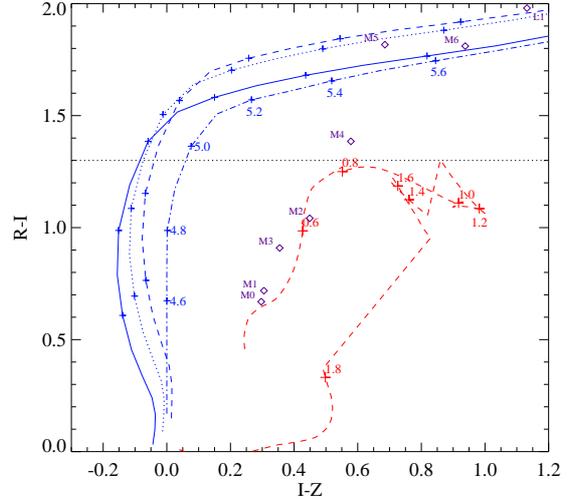}
\caption{Colour modelling of high redshift star forming galaxies (blue) using a variety of stellar ages. A simple SED flat in f$_\nu$ is represented by the dash-dot line. The other three colour tracks are star forming galaxies from \citet{maraston05} with increasing stellar age of 10 Myrs (solid line), 100 Myrs (dotted line) and 400 Myrs (dashed line). All have had appropriate IGM absorption applied as described in the text.  Also shown are intermediate redshift elliptical galaxies (red) from \citet{maraston05} models, assuming a formation redshift of z=5 and a star formation e-folding time of $\tau$=0.5Gyr, and cool, low mass stars with labels indicating subtype (purple) from \citet{hawley02} templates, using appropriate FORS2 filter set for this study. Redshifts are noted for the high and intermediate redshift galaxies.}
\label{fig:riz}
\end{center}
\end{figure}

Figure \ref{fig:riz} shows the expected optical colour evolution of 
a high redshift LBG represented 
by the blue solid, dotted and dashed lines labelled from redshifts of 4.6 to 5.6.
The dot-dash line represents a SED which is flat in $f_\nu$ and the solid, dashed and dotted blue
lines are star forming galaxies taken from \citet{maraston05} integrated galaxy spectral synthesis models
(updated PEGASE models \citep{fioc97} of the same age produce similar colours). Each track represents a 
different stellar age, 10 Myrs (solid), 100 Myrs (dotted) and 400 Myrs (dashed), the oldest reasonable stellar age
given the redshift range probed.

All models have been adjusted for the effects of absorption by intervening 
hydrogen clouds in the intergalactic medium (IGM). We use the models of \citet{madau99}
 for the evolution in the number density and column density distribution of hydrogen clouds
in the IGM to estimate the absorption as a function of redshift due to the Lyman-series 
and Lyman limit. In common with other authors \citep[e.g.][]{bouwens07}, we modify
the original Madau models to account for the higher absorption seen in the spectra of high
redshift quasars \citep{songaila04} by requiring a more rapid evolution in the number 
density of absorbing clouds ({\it i.e.} (1+z)$^3$ rather than (1+z)$^{2.46}$) beyond z=4.5.

As the stellar
age increases, the colours of the galaxies become redder. Although models
with varying extinctions are not shown here to maintain the clarity of figure \ref{fig:riz},
there is a similar reddening
effect with increasing extinctions. Applying the \citet{calzetti00} dust model to the 
galaxy models shown, the $R-I$ colour reddens by $\sim$0.2 magnitudes and the $I-z$ colour 
by $\sim$0.1 magnitudes per 0.1 increment in E(B-V) from 0 to 0.5. Given the single $R-I$ 
colour cut used in this work we are able to include any $I$-band selected candidate regardless
of it's intrinsic continuum slope and corresponding $I-z$ colour. Results from the spectroscopic
follow-up suggest that the majority of $z\sim5$ LBGs have intrinsically blue colours in the observed optical bands at wavelengths shorter than the Balmer break with little
reddening, despite the fact that much redder objects could have been observed, justifying the use
of relatively blue galaxy models in understanding the optical selection function. While
older, redder objects do exist at $z\sim5$ \citep[e.g.][]{verma07}, they are a minority population.

The single colour selection used here could have
introduced a high number of candidates which are only detected in a single band potentially
leading to a larger number of spurious sources caused by noise spikes. Our spectroscopy shows that 
this is not the case. Within the
photometric selection there were 40 objects 
with a single band detection in the $I$-band. Of these,
15 were observed spectroscopically down to the faintest $I$-band magnitudes
and only one was not detected in the continua of the
subsequent spectra. In the complete spectroscopic program \citep{douglas09}, which targeted
fainter and bluer objects, an additional 10 objects were only detected in the
$I$-band to $I<26.3$ and all of which had continuum spectroscopically detected
and are therefore real objects.

These spectroscopic results suggest only one of 25 observed sources could be spurious
giving an upper limit of 4\% of the single band detections for the fraction of spurious sources. 
At these faint magnitudes, a lack of spectral flux
is not conclusive evidence that such a source is spurious. Thus, this upper limit
could be a considerable overestimate. Assuming the spectroscopic results are representative there are,
at most, only 2 objects within the sample predicted to be spurious and as such, this 
is not treated as a significant problem. 

The number of spurious sources was also checked by performing identical source detection routines 
on inverted images. No spurious sources were identified using this method.

\subsection{Common Red Interlopers}
\label{sec:interlopers}

While a simple two or three-band colour cut can efficiently select a
sample of distant LBGs, such a sample will inevitably include
 low mass Galactic stars and intermediate redshift passively
evolving or reddened galaxies \citep{stanway08b} even without photometric scatter. By using the
additional near-IR photometry we can
substantially decrease the number of these interlopers in our final
sample.

\begin{figure}
\begin{center}
\includegraphics[width=8cm]{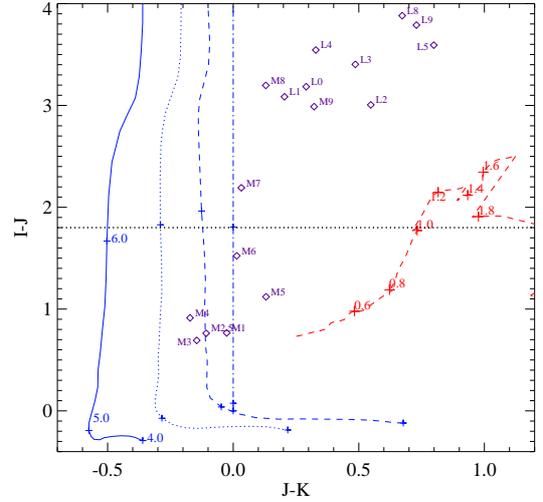}
\caption{Colour modelling high redshift star forming galaxies, intermediate redshift ellipticals and low mass stars as described in figure \ref{fig:riz}. The dotted line represents the 2$\sigma$ colour limit of the $I-J$ colour at the flux limit of the sample.}
\label{fig:ijk}
\end{center}
\end{figure}

The first major potential interlopers that we need to consider  
are cool, low mass Galactic stars. To estimate accurate colours for the 
stellar interlopers applicable
to the specific filter set used here, the stellar templates derived by
\citet{hawley02} were processed through the FORS2 filter set
including detector response. Figure \ref{fig:riz} shows the $R-I$ and
$I-z$ colours for M-type dwarfs. It shows that late M-types will be
included in the sample with a $R-I>1.35$. Figure \ref{fig:ijk} shows
\citet{knapp04} near-IR stellar spectra convolved with the
appropriate filter set. It shows that the $J-K_{s}$ colour is relatively
flat for M- and L-type dwarfs, however a larger range is observed in
the $I-J$ colour spanning from 0.8 for the early M-type dwarfs to 3.5 for
the late L-type dwarfs.  

The 2$-\sigma$ magnitude limits for our
near-IR data are typically $J<24.5$ and $K_{s}<23.7$. Given these near
infrared depths late M-type and L-type dwarfs should be excluded
from the high redshift sample through their $I-J$ and $J-K_{s}$ colours with
limits of $I-J>1.8$ (shown by the black dotted line in figure \ref{fig:ijk}) 
and $J-K>0.8$ for near-IR detections. Although early-type M-stars are
not red enough to be detected in the near infrared imaging at the flux limit
of the sample, their
relatively blue optical colours ($R-I\leq1$) would exclude them from
the high redshift selection.  By contrast, true LBGs are expected to have 
relatively flat colours with $I-J<1.8$ and $J-K<0$ \citep{verma07} and as such
should not be detected in the near-IR imaging. As we note later, an advantage of
using a single $R-I$ colour cut is that it includes intrinsically red objects. However,
as noted previously,
the majority of spectroscopically confirmed high redshift galaxies had relatively 
blue continuum, supporting the use of the comparatively blue models.

By incorporating the redder near-IR wavebands into the selection criteria
 we can exclude late M-type and L-type dwarf stars and early M-types from M1 to M3. The only 
stellar populations which have colours that can not be isolated using 
this imaging data set are M4 to M6 dwarf stars. However, the number density of M-stars
is expected to decrease with fainter magnitudes.

The other major potential interloper of high redshift samples
 are the class of intermediate
redshift elliptical galaxies often referred to as EROs (extremely red
objects). Such interlopers have a redshift between 0.5 and 2.0 and
have redder optical to near-IR colours than the high redshift star
bursting galaxies. Figure \ref{fig:riz} shows the $RIz$ colour
evolution for a passively evolving intermediate redshift source.
 This population of galaxies could enter the
selection around a redshift of $z=0.8-1.6$, where the 4000\AA\ break
falls in the $I$-band. When combined with the underlying red colour of
a passively evolving stellar population, this can lead to intermediate
redshift objects with sufficiently red $R-I$ colours falling into the
simple colour selection.  However, these objects can again be identified 
and removed from the high redshift sample using near-IR imaging of sufficient depth. 
Different levels of dust could also redden these intermediate redshift
galaxies in the optical encroaching them into the high redshift selection.
Due to the $I$-band selection, dust reddening would lead to bright
$z$, $J$ and $K_{s}$ colours leading to such galaxies being rejected from the sample by the near-IR
criteria in sub-section \ref{sec:nearir}.

The intermediate redshift galaxy model from figure \ref{fig:riz} is plotted
in figure \ref{fig:ijk} which illustrates the $I-J$ and $J-K_{s}$ colour
evolution. The model indicates that at a redshift of $z\gtrsim1.0$ the
galaxies would be detected in the near-infrared imaging with an $I-J$ colour redder than
the limits of the imaging even at the flux limit of $I=26.3$. Intermediate
galaxies at lower redshifts ($z<1$) are in principle removed by the 
initial optical selection, however stochastic photometric uncertainties may cause 
them to reenter the sample.

Although the near-infrared data available to this study is not deep
enough to detect the individual high redshift candidates, the colours calculated
above imply that many lower redshift interlopers (early and late M-type and L-type 
dwarf stars and ellipticals at $z>1$) are
detectable and can be identified. As such, any candidate with a clear near-infrared
detection was removed from the high redshift galaxy 
sample. As figure \ref{fig:ijk} shows, galaxies at $z\sim5$ are not expected 
to be detected in the near-IR imaging. Only galaxies with a high extinction, 
reddening the optical to near-IR colours, could expect to be detected. For the
mean $I$-band magnitude of the sample, two magnitudes of reddening in the $K_{s}$-band would
be required for the object to be detected in the $K_{s}$-band imaging. 
Studies such as those of \citet{verma07}, which model the spectral energy distribution 
of similar high redshift galaxies, expect relatively low extinctions making
such dusty objects rare. From Monte Carlo simulations using the probability distributions
of models rather than the formal best fitting scenario, \citet{verma07} found no objects
with reddening greater than one magnitude and less than 10\% of the formal best fitting
models had sufficient reddening to be detected the ERGS $K_{s}$-band imaging.
Hence, removing all objects with a near-IR detection
will reduce the interloper fraction while not significantly effecting the completeness of the 
high redshift sources given the rarity of sufficiently dusty sources.

\subsection{Near-IR selection criteria}
\label{sec:nearir}

\begin{figure}
\begin{center}
\includegraphics[width=8cm]{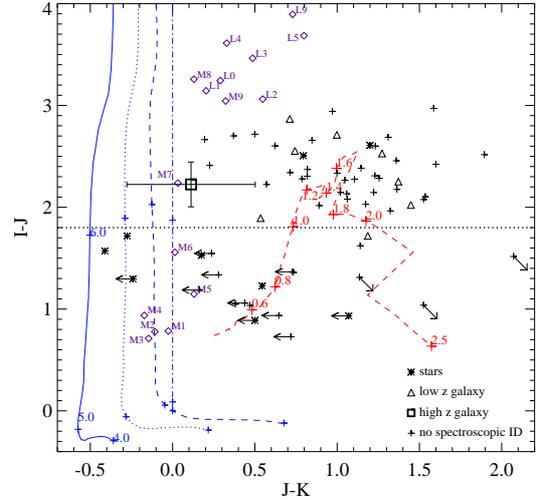}
\caption{Colour modelling high redshift star forming galaxies, intermediate redshift ellipticals and low mass stars as described in figure \ref{fig:riz}. Overplotted are the colours of sources detected in the near-IR imaging and subsequently rejected from the high redshift sample. The colour error bars are shown for the confirmed high redshift galaxy.}
\label{fig:ijk3}
\end{center}
\end{figure}

Using the near-IR imaging to identify common interlopers we exclude objects with 
$J< 24.5$ and/or $K_{s}<23.7$ from the LBG sample. At close to the $I$-band limit
these near-IR limits indicate colours incompatible with a high redshift identification
(figure \ref{fig:ijk3}). The effect of the near-IR colour
cut is to exclude 66 objects. Within this sample 63 objects had $I-J>0.7$
and $J-K>0.1$ separating them from the theoretical colours of low extinction high redshift
galaxies. The three remaining objects had blue $J-K$ colours approaching the
high redshift colour tracks. However, all these objects were bright ($I<24.9$) and 
unresolved in the HST imaging making them likely contaminants. Subsequently, all
three were spectroscopically confirmed to be at low redshift \citep{douglas09}.

During our programme of spectroscopic follow-up we observed a
representative sample of rejected objects in order to confirm that these
objects had a low probability of being genuine $z\sim5$ LBGs. Of the 66 sources with
near-IR detections, 24 
were observed spectroscopically. Six spectra contained flux which was too faint to identify, 
one source was confirmed to be at high redshift (a continuum break galaxy at $z=5.45$) and 17 were confirmed to be galactic stars or intermediate redshift galaxies (figure \ref{fig:ijk3}). 95\% of all sources detected in the near-IR imaging with sufficient spectral signal-to-noise were confirmed to be at low or intermediate redshift. 

\begin{figure}
\begin{center}
\includegraphics[width=8cm]{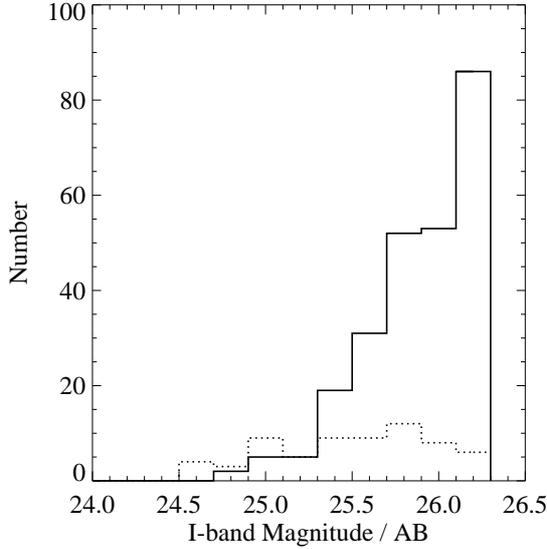}
\caption{Distribution of observed $I$-band magnitudes of final photometric sample (solid line) with near-IR detected sources rejected and the sources removed from the sample because of near-IR detections (dotted line). }
\label{fig:ihist}
\end{center}
\end{figure}

Although the LBG sample covers a relatively small range in $I$-band magnitudes, the interloper fraction
cannot be assumed to be constant with flux. The $I$-band magnitude distribution of the LBG candidates
and the near-IR detected objects shown in figure \ref{fig:ihist} demonstrates the varying interloper 
fraction as a function of magnitude. Although it is likely that some faint interlopers may not been 
identified by the near-IR data given the depths of the images (see section \ref{sec:contamfrac}), the flat distribution observed suggests that the interlopers have the biggest statistical effect at
bright magnitudes. This is at least in 
part due to the nature of the interloping samples. The fraction of stellar objects
 is expected to decrease at fainter magnitudes ($I>26$) compared to 
brighter magnitudes due to the distribution of M-type dwarfs in the Galactic halo. 
This is also seen in the low level of contamination by stars in faint samples of
HST-selected LBG candidates \citep{bunker04}. Similarly, at these
faint magnitudes the intermediate redshift elliptical galaxies
originate from the flatter faint-end of their luminosity function
rather than the comparatively steeper exponential bright-end, assuming
a typical Schechter function fit to the $z=1$ luminosity
function. With the candidate high redshift galaxy number counts
increasing towards fainter magnitudes this results in a lower fraction
of intermediate redshift galaxies at $I>26$.

\section{The $z\sim5$ Galaxy Candidate Sample}
\label{sec:highz}

The final photometric selection consisted of 253 objects in the
refined sample, all with multiband photometry consistent
with a redshift of $z\sim 5$ or higher, using the
strict criteria discussed above. Figure \ref{fig:ihist} shows the 
observed $I$-band flux distribution for the refined sample of 253
objects and the 66 removed because of their near-IR detections. 

Significant field-to-field variation in the number of high redshift candidates
was observed as shown in table \ref{fieldtofield}. The range of number
densities observed for the refined sample in individual fields can not
be explained by cosmic variance when using the estimator of \citet{trenti08}
suggesting observations of truly under and overdense fields. In contrast,
the variation in the number density of near-IR detected sources can be explained by
cosmic variance, estimated by \citet{trenti08} assuming a lower redshift (z=1), 
with the exception of only two fields.

\begin{table}
\caption{Source density per arcmin$^2$ for each survey field.}
\label{fieldtofield}
\begin{center}
  \begin{tabular}{|c|c|c|}
\hline
\bf Field& Refined sample & Near-IR sources\\ \hline
J1037.9-1243 & 0.47 & 0.23 \\ 
J1040.7-1155 & 0.85 & 0.26 \\ 
J1054.4-1146 & 0.26 & 0.22 \\ 
J1054.7-1245 & 2.09 & 0.19 \\ 
J1103.7-1245 & 0.67 & 0.18 \\ 
J1122.9-1136 & 0.65 & 0.52 \\ 
J1138.2-1133 & 0.64 & 0.07 \\ 
J1216.8-1201 & 1.55 & 0.26 \\ 
J1227.9-1138 & 0.78 & 0.21 \\ 
J1354.2-1230 & 1.29 & 0.25 \\ \hline

\end{tabular}
\end{center}
\end{table}

\subsection{Single versus two colour cut}

Due to the shapes of the filter bandpasses and the typical colours of low mass stars,
a two colour selection, {\it{i.e.}} criteria in $R-I$ and $I-z$ colours, is unable 
to efficiently select high redshift galaxies while eliminating M-type dwarf
stars. Given this and the range of $I-z$ colours predicted for high redshift
objects, a single $R-I$ criteria was adopted. While the absence of an $I-z$ colour cut may lead to
a higher contamination fraction in a purely optically selected sample, the
addition of near-IR selection criteria minimise this effect as discussed in section \ref{sec:select}.

\begin{figure}
\begin{center}
\includegraphics[width=8cm]{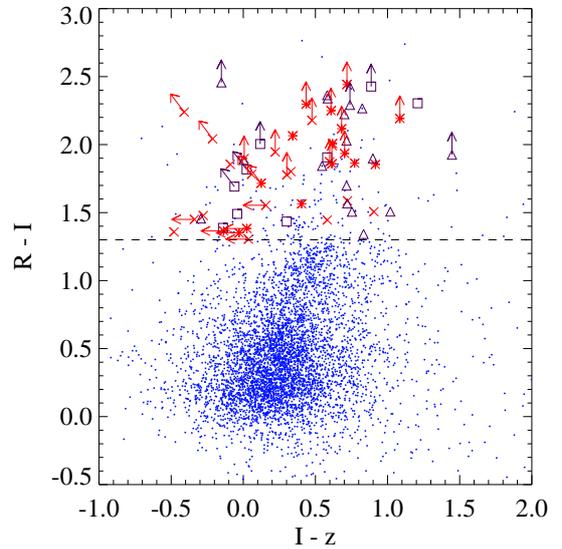}
\caption{Colour-colour diagram showing spectroscopically confirmed high redshift galaxies (crosses - line emitters, asterisks - continuum breaks) and confirmed lower redshift objects (galaxies and stars where squares are objects undetected in nIR imaging and triangles are objects detected in nIR) compared to the field populations (points). Arrows represent 2$\sigma$ colour limits when there is no detection in corresponding band.}
\label{fig:1216riz}
\end{center}
\end{figure}

The $R-I$ and $I-z$ colours of the spectroscopically observed sources within the photometric sample
are shown in figure \ref{fig:1216riz}.
The small dots show the range of colours seen in the catalogue of all objects in a single
field. Over-plotted are the objects which were spectroscopically
confirmed to be at high redshift (asterisks are galaxies identified
using only their continuum break and crosses are galaxies identified
with Ly$\alpha$ line emission) and confirmed lower redshift
galaxies and stars with and without near-IR detections (triangles and squares respectively). 
A number of sources, spectroscopically confirmed to be low redshift galaxies or galactic stars, have $I-z$
colours bluer than that expected from the models in figure \ref{fig:riz}. This is most likely caused by a combination
of photometric errors and different intrinsic properties of the sources to that assumed in the models. Figure \ref{fig:1216riz} illustrates that without the near-IR imaging many low redshift objects
would not be distinguishable using only the optical colours. Also, had an 
$I-z$ colour criteria been applied to eliminate lower redshift contaminants, many
high redshift galaxies would have been excluded from the sample.

A variety of $R-I$ and $I-z$ colour cuts could be used to isolate the stellar 
interlopers from the LBG candidate sample, {\it{e.g.}} $I-z<0.5$. Adopting
this additional colour cut should, 
in principle, exclude any stellar interlopers, however for the case of $I-z<0.5$, 36\% of the spectroscopically confirmed
high redshift sources would be rejected from the sample and only three 
of the nine spectroscopically confirmed lower redshift interlopers 
would be removed. Thus, the application of such a colour cut
for this combination of filters would significantly reduce the number of high redshift candidates 
 biasing the sample to blue rest-frame UV continuum, while not significantly reducing the relative contamination.

The additional near-IR selection criteria is an effective substitute for an $I-z$ colour cut, removing
common interlopers without influencing the high redshift LBG sample.

\subsection{Contamination Fraction from Spectroscopic Observations}
\label{sec:contamfrac}

It is possible that for the
optically-fainter objects we fail to reject contaminants because they
are too faint to be detected in the near-IR data. At the flux limit of
the sample there is a colour limit of $I-J<1.8$ and $I-K_{s}<2.6$
which does not exclude all possible low redshift contaminants even in the
absence of photometric scatter. Lower redshift
objects can also scatter into photometric selections through photometric error
and extreme colours.

\begin{figure}
\begin{center}
\includegraphics[width=8cm]{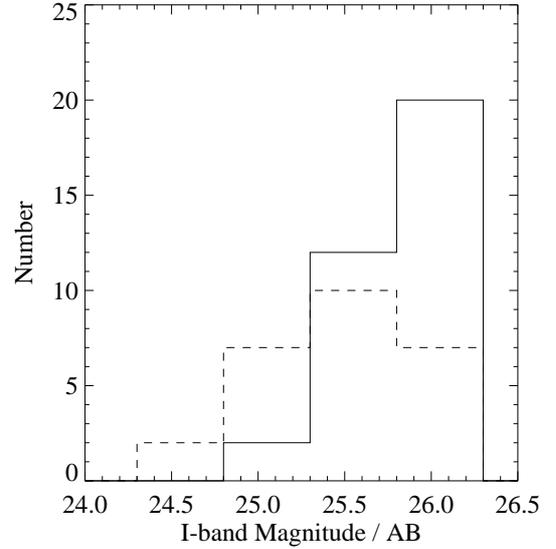}
\caption{Distribution of $I$-band magnitudes for spectroscopically confirmed high
redshift galaxies (solid line) and lower redshift objects (galaxies and stars, dashed line) within the raw photometric sample.}
\label{fig:spechist}
\end{center}
\end{figure}

Our spectroscopic programme was optimised to maximise the number of
spectroscopically-confirmed $z\sim 5$ objects and had sufficient slits to
sample robustly all objects which obeyed the optical selection. Consequently 
we were able to test if a given selection criteria improved the sample.
A third of all photometric candidates with
a range of $I$-band flux, colours and near-IR detections were observed.
The spectroscopic identifications of this subsample can be used to estimate 
the fraction of contamination in the refined photometric high redshift sample.
Although a spectroscopic identification is harder to achieve for intrinsically
fainter objects, figure \ref{fig:spechist} shows that our confirmed high redshift LBGS are not biased
to brighter $I$-band magnitudes and have a distribution similar to that of the
refined sample as a whole (figure \ref{fig:ihist}). Contrary to what would 
normally be expected, fainter objects can be
 identified spectroscopically if they have a strong Ly$\alpha$ emission line
or a clear continuum break in a comparatively noise free region of the spectrum. 
These features can often make high redshift galaxies easier to identify than
lower redshift objects without any emission lines. At low signal-to-noise, lower
redshift objects can show essentially featureless continuum making any redshift 
determination difficult, if not impossible. As such, any observed objects which can 
not be identified in their spectra due
to low signal-to-noise, often up to 50\% of the observed sample, cannot be 
assumed to lie at high or low redshift; either scenario is possible.

Of the 253 objects in the refined sample, 83 were observed on the spectroscopic
masks. Within this subsample 33 objects were confirmed to lie at $z>4.6$, 9 were 
identified as intermediate galaxies or stars and 41 were detected spectroscopically
but the signal-to-noise was too low to allocate a firm identification. Of the identified
objects, 20\% were at a redshift lower than targeted. Using the $K_{s}$-band imaging,
the fraction of high and low redshift objects in the unidentified sample is estimated.

While none of the individually confirmed high redshift galaxies were detected in $K_{s}$, by
stacking 30 images we determined that their average colour was
$I-K_{s}<0.0$. Similarly, a typical colour for the confirmed low redshift sources and
the unidentified objects was obtained. The nine confirmed low redshift sources
gave colour of $I-K_{s}=1.3$. 201 unidentified objects in the refined sample, including those
observed spectroscopically but resulted in low signal-to-noise, had an average colour of 
$I-K_{s}=0.3$. Where there was no detection in the stacked $K_{s}$-band image the 2$\sigma$
limit for a 2$''$ circular aperture was measured using background regions. Given the need
for adequate background regions surrounding the central stack to estimate the depth of the image, a small 
number of sources were not included as they were positioned close to the edge of
the near-IR field.

Assuming that the unidentified sample is a combination of the high ($I-K_{s}<0.0$) and low redshift ($I-K_{s}=1.3$)
populations, the typical colours imply the unidentified sample contains at least $\sim$75\% high redshift objects 
and less than $\sim$25\% low redshift objects. Combining these results with the spectroscopic
identifications gives a contamination fraction of $\sim$20\% for the refined sample. Applying the
same technique to all optically selected objects including the near-IR detected objects, the contamination fraction is estimated to be around $\sim$40\%. The unidentified near-IR detected sources have an average colour of $I-K{s}=3.2$ (compared with a colour of $I-K{s}=2.6$ for the near-IR detected objects confirmed to be at low redshift) which implies a very high level of interloper within this subsample. The higher contamination rate estimated for the full sample clearly demonstrates the utility of the near-IR rejection of interlopers in addition to the purely optical selection.

There have been several other spectroscopic studies of $z\sim5$ LBGs, for example
\citet{ando07} and \citet{vanzella09}. The differences in selection function and
filter sets used in these studies makes a direct comparison of the level of contamination
difficult. One feature that all high redshift spectroscopic surveys share is the large fraction of
unidentified sources due to insufficient signal-to-noise in the spectroscopy. This can be
up to 50\% of the observed sample. Such studies can appear reliable because most or all of the
spectroscopically identified objects are at high redshift, {\it{e.g.}} \citet{ando07} did not
identify any lower redshift objects spectroscopically. However, care must be taken when assuming
the same contamination rates for the spectroscopically unidentified sources and subsequently 
for the whole photometric sample, especially if only the brightest candidates have been targeted.

Although it is difficult to compare contamination fractions of different samples due
to different selection, filters and instrument combinations \citep{stanway08b}, it 
seems that typical contamination rates are between 10\% and 20\%. \citet{stanway08b} 
shows that with contamination fractions of this level, measurements such as the angular
correlation function of photometrically-selected LBG samples can be
compromised because the expected contaminating population can
themselves have a clear correlation signal. Also, the difference in
optical-to-near-IR colour between contaminants and confirmed high redshift LBGs
demonstrated here can also cause problems in estimating the typical
multi-wavelength spectral energy distributions of LBGs. If a
population of contaminants individually-undetected in the near-IR are
combined in a stacking analysis with true LBGs, they will redden the
resulting SED, and artificially inflate the estimated age of the
stellar population in a typical LBG or dust fraction.

\section{Completeness Determinations}
\label{sec:completeness}

The completeness of the selection in each field was estimated by
carrying out two kinds of simulations. The first is a simulation of
recoverability which depends on the quality and depth of the imaging.
This involved injecting artificial sources with known
properties into our photometry data in order to determine the
efficiency of SExtractor in recovering them. The second simulation investigated 
the effects of the selection criteria and image quality on a model
spectrum.

\subsection{Recoverability}

For each field, 300,000 artificial galaxies were
generated covering a range of magnitudes from 22.0 to 28.0 and 
a $R-I$ colour was assigned within the range 0.0
to 3.0. A surface brightness profile was chosen
from a sample of four stellar profiles measured directly from the images to
reflect the seeing conditions (the high redshift candidates are
unresolved in ground-based data),
and a selection of HST/ACS images of two or more galaxies seen at small angular separations
selected in the GOODS data of the Chandra Deep Field South (CDFS). The
high resolution multiple systems were convolved with a Gaussian
profile to match the ground-based seeing conditions and were chosen
with the same frequency as seen in the CDFS ($\sim$20\%) to mimic this
population. Using the selected magnitude, colour and profile, a `simulated'
galaxy was created for the $I$-band and $R$-band images and placed at
random coordinates within the field excluding previously masked areas
near bright objects used in the photometric selection. By placing these galaxies
directly onto the science image, the noise properties of the image
were added to the simulated galaxy. The process was repeated one
thousand times for each field as no more than 300 simulated galaxies were
placed on any one image to prevent mutual overcrowding affecting the
results.

Each new image with the simulated galaxies was processed through
SExtractor in the same way as the original science data, creating
catalogues containing the simulated galaxies as observed on the images. After these were matched with
the original galaxy information on magnitude, colour and position, a
catalogue was created containing the known simulated magnitudes and
the recovered SExtractor magnitudes of each fake galaxy. 

\begin{figure}
\begin{center}
\includegraphics[width=8cm]{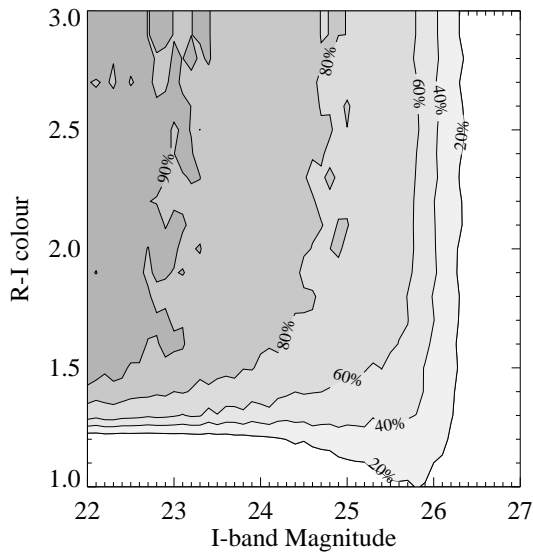}
\caption{Contours illustrating the success rate of recovering galaxies within the high redshift colour selection. The levels shown are 20\%, 40\%, 60\%, 80\% and 90\%. The x-axis is the original $I$-band magnitude of the `fake' galaxy inserted into the images and the y-axis is the original colour assigned to this `fake' galaxy. The contours show the percentage recovery rate of these `fake' galaxies using the same method as the high redshift selection. It is shown that the completeness of object
recovery is generally independent of colour with the results
reflecting the increasing photometric errors at fainter magnitudes.}
\label{fig:comp}
\end{center}
\end{figure}

The high redshift galaxy candidates were selected on two main
criteria, a magnitude limit and a colour cut. The recoverability of 
the galaxies using the high redshift selection of $I<26.3$ and $R-I>1.3$ 
was measured for a range of
colours and magnitudes and is shown in the contour plots of
figure \ref{fig:comp}. Eight of the ten fields show comparable
results, with a completeness of 25\% at the $I$-band flux
limit. However two fields had consistently lower completeness which
was caused by the two worst seeing conditions for the $I$-band
observations, 0.73$''$ compared to an average of 0.6$''$ over the
other eight fields. This spread the received flux over a larger area
causing the objects to drop below the surface brightness limit. If the
objects were detected, the photometry, made fainter by the seeing
conditions, is corrected by the appropriate aperture
corrections. Figure \ref{fig:comp} shows an average of the ten survey
fields, with the simulated $I$-band magnitude and $R-I$ colour along
the x and y axes respectively. The contours are placed at 20\%, 40\%,
60\%, 80\% and 90\% of inserted galaxies recovered, using the colour
selection of $I<26.3$ and $R-I>1.3$. Figure \ref{fig:comp} shows that
the completeness of object recovery is generally independent of colour
with the results reflecting the increasing photometric errors at
fainter magnitudes. The maximum completeness never rises to more than
$\sim95\%$ due to object confusion despite areas near bright objects being masked.

Figure \ref{fig:comp} implies that only 60\% of galaxies with an
intrinsic $R-I$ colour of 1.3 would be included in the sample. This 
is not surprising as only $\sim$50\% of objects would be expected to 
be included in the sample at the colour cut. Even a small photometric
error in the $R$ or $I$-band would cause the colour to become too blue.
The applied $R-I$ colour criterion was chosen from the previous work of \citet{lehnert03}
who found a colour cut of $R-I>1.5$ selects galaxies at $z>4.8$ using the same FORS2
filters used in this work and models from \citet{fioc97}. By adopting a bluer
colour cut, completeness at this selection colour is much higher. Although this
could introduce a higher contamination fraction, the additional near-IR imaging
should minimise this.

Also, the objects that were
recovered with intrinsic colours bluer than our cut represent a
population of objects that can contaminate the sample, rather than
one that affects the sample completeness. These objects have their
measured colours reddened by the statistical errors on the
photometry. At $I=26$ a substantial fraction ($\sim 20 \%$) of objects
with intrinsic $R-I=1.1$ can be included in a sample with an observed
cut of $R-I>1.3$ given the photometric properties of our source
data. Such objects are likely to be at $z<4$ and could be a
substantial contaminating population depending upon their luminosity
function. In reality, such objects would often have significant $V$-band flux 
or near-IR detections which would rule them out of the high redshift candidate
sample.

\subsection{Effect of Selection Criteria}

\begin{figure}
\begin{center}
\includegraphics[width=8cm]{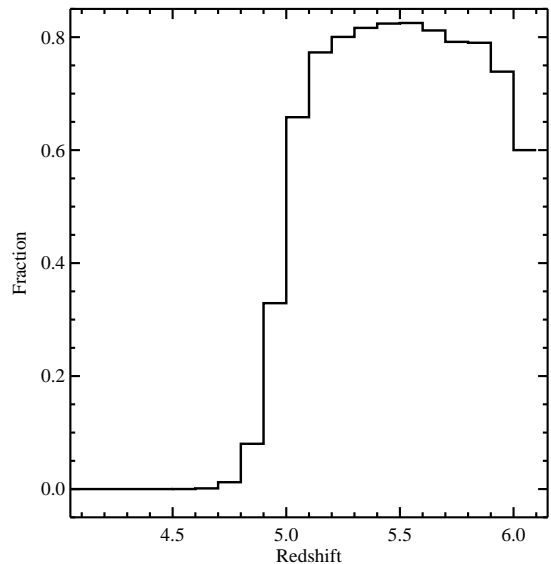}
\caption{The fraction of galaxies recovered as a function of redshift with an intrinsic $I<26.3$, $R-I>1.3$ and $V>27.5$ given the photometric errors of the imaging data. The luminosity function used to generate the predicted number counts was that measured in \citet{bouwens07} for $z\sim5$.}
\label{fig:frac}
\end{center}
\end{figure}

The second simulation determined the completeness of our final sample
given the interplay of photometric cuts used to reject contaminants
and the statistics of the photometry data, as a function of
redshift. An artificial catalogue of $10^8$ sources was generated with a
$I$-band magnitude distribution determined by the $z=5$ \citet{bouwens07} luminosity
function. Each galaxy was assigned a redshift (in 0.01 redshift bins) which determined 
its $I$-band magnitude and $R-I$ colour assuming a flat spectrum in $F_{\nu}$
and accounting for IGM absorption using the modified Madau perscription as a function
of redshift (described in section \ref{sec:optical}). The $V$-, $R$- and $I$-band magnitudes of the
artificial sources were then independently perturbed to
match the typical noise properties of the imaging data and the typical weak lensing
characteristics from the foreground clusters. The fraction
of objects which were then recovered using the previous $z\sim5$
selection criteria ($I<26.3$, $R-I>1.3$ and $V>27.5$) as a function of redshift is shown in
figure \ref{fig:frac}. In total, around 80\% of galaxies at
$5.0<z<6.0$ are selected using the photometric colour criteria of this
work but completeness falls off rapidly below $z=5$.

\section{Morphology}
\label{sec:morph}

The morphologies of the high redshift candidates were studied using
high resolution images from the Hubble Space Telescopes using the
ACS camera and F814W filter which cover nine out of the ten 
survey fields. Each field was observed for 1 orbit with
the central regions observed for a total of 5 orbits. 

Catalogues from the HST/ACS images were created using SExtractor in
order to obtain half-light radii for the potential high redshift
galaxies. Only sources with the deepest, 5 orbit depth of data are
discussed.

\begin{figure}
\begin{center}
\includegraphics[width=8cm]{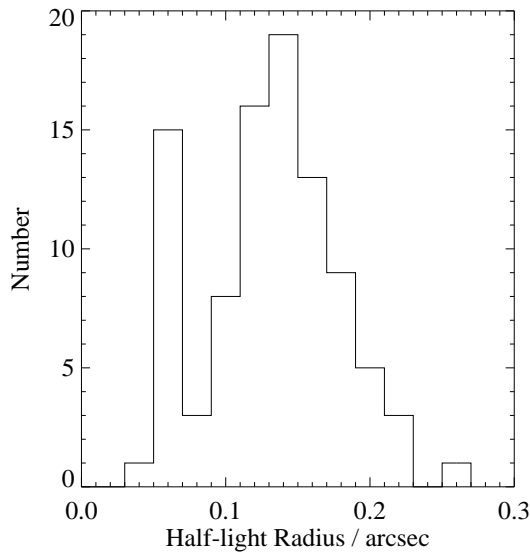}
\caption{Distribution of half-light radii, in arcseconds, of the refined high redshift sample with the deepest HST data. Radii below 0.07$''$ are unresolved.}
\label{fig:morphrad}
\end{center}
\end{figure}

The majority of high redshift candidates in the refined sample were found to be
resolved with individual peaky components. They have half-light radii between 0.1$''$ and 0.26$''$ (figure \ref{fig:morphrad}) which
corresponds to a range of 0.3-1.6 kpc at a redshift of $z\sim5$, with
a mean value of 0.14$''$ or 0.8 kpc in agreement with
\citet{bremer04}, the purely space-based sample of \citet{conselice09} and the higher redshift galaxies of \citet{bouwens06}. A subset have formally unresolved radii ($\sim$0.06$''$).
However, these objects are generally faint with limited S/N that could
be effecting the measurement and fitting procedures. While
some of these could potentially be stars, their colours do not support
this hypothesis. It is also possible to select AGN such as that described 
in \citet{douglas07} which could be unresolved in the HST imaging.

\begin{figure}
\begin{center}
\includegraphics[width=8.5cm]{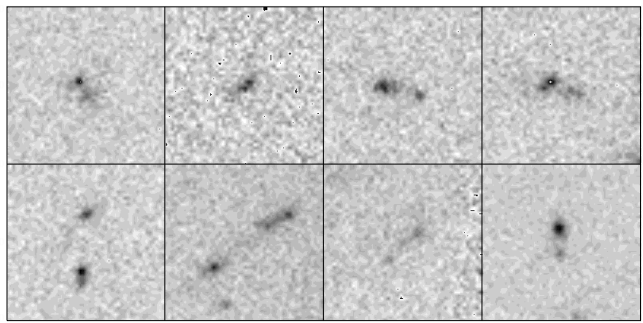}
\caption{Examples of high redshift galaxy candidates with multiple components or disturbed morphology visible in the high resolution HST/ACS F814W imaging. Each image segment is $2.5''\times2.5''$.}
\label{fig:morph}
\end{center}
\end{figure}

Over the 9 fields, 31 out of 95 (33\%) candidates with deep
HST data show disturbed or asymmetric morphology or are part of a multiple system on
scales from 0.2$''$ to 1.2$''$ or 1.3kpc to 7.5kpc.  Examples of multiple
systems, which are not resolved in ground based imaging, are shown in figure \ref{fig:morph}.
Each individual component has a half-light radius consistent with the the isolated
objects. It is not clear whether these are most likely to be multiple interacting systems or
UV-bright knots in an underlying larger system. 

Of the multiple systems, 10 candidates are close pairs and 4 are part of a larger
collection of sources with a mean separation of 0.5$''$. Objects with separation distances of more
than 0.8$''$ are barely resolved in the ground-based imaging and
extend with the same orientation. The fractions of objects observed with multiple members (15\%)
agrees with the work of \citet{conselice09} in the Hubble Ultra Deep Field (HUDF) where $\sim$19\% of their 
V-drop sample were found to be in pairs. With this much deeper data, the possibility that the observed
galaxy pairs are a larger object with UV bright peaks is less. As the sample fractions showing galaxies
pairs for the deep HUDF sample and the sample presented here are similar, it is likely that these are the
true pair fractions for each sample.

The 17 candidates which show disturbed or asymmetric morphology were identified visually. Even in the deepest available HST imaging the objects are small and faint making any detailed measurement of their structure
challenging. For a similar sample in the HUDF, \citet{conselice09} observed the same fraction exhibiting asymmetric
morphologies. With the much deeper observations of the HUDF, more precise measurements of the structure
and potential asymmetry of the high redshift galaxy candidates was possible. Although only visual classification was used for the ERGS sample, the agreement of asymmetric sample fractions suggest that such a classification is reliable. It also suggests that the fraction of multiple of asymmetric systems does not change with UV luminosity over the range probed by the two studies.

\section{The UV luminosity function}
\label{sec:nc}

One apparently straightforward measurement of the properties of $z\sim
5$ LBGs that can be made with a photometric sample is the rest-frame
observed UV luminosity function. For this to be reliable and meaningful, the
 contamination and completeness of the sample must
be understood. While our reliable photometric sample is constrained to
bright magnitudes in comparison to some published data
\citep[e.g.][]{bouwens07}, it can be used to explore the brighter end
of the luminosity function.

In order to calculate a luminosity function which represents the observed
luminosity distribution of galaxies, the following method, which is essentially 
the same as that used by \citet{bouwens07}, was used. Monte carlo simulations
were used to create artifical populations of LBGs across the redshift interval
probed photometrically, with corrections for photometric errors, completeness 
and weak lensing. After an identical photometric selection, the resulting
artifical LBG candidates represents what would be observed given the
intial luminosity function assumed. When compared to the real, observed LBG 
candidates, a best-fitting intrinsic luminosity funtion can be found.

A range of Schechter luminosity functions (defined by the characteristic magnitude $M^*$, 
the faint end slope $\alpha$ and normalisation $\phi^*$) were generated across a broad
spectrum of parameters, from $-19.65$ to $-21.45$ in $M^{*}_{UV}$,
 faint end slopes, $\alpha$, between $-1.40$ and $-1.95$ and $\phi^*$
fitted freely, and were compared to the observed number counts of the $z\sim5$
galaxy candidates.

For each proposed luminosity function the following steps were used to
create a catalogue of galaxies which reflects the assumed luminosity
function and the properties of the observed fields. Within 0.01 redshift bins
from $z=4$ to $z=6.1$, the luminosity distance, luminosity function and IGM absorption were used 
to determine the apparent magnitude distribution and number density of objects scaled according to the 
volume within the redshift slice (per arcmin$^2$). The colours of the objects
in this redshift bin were calculated according to the FORS2 instrumental response
and filter profiles assuming the galaxies were flat in $F_\nu$ and incorporating
\citet{madau99} IGM absorption with a higher density of absorbing systems at high redshift as discussed 
in section \ref{sec:optical}. The galaxy catalogues for each redshift bin were combined
to create a sample with a magnitude, colour and redshift distribution which reflects the
proposed luminosity function. 

To simulate the properties of the survey fields, each galaxy in the artificial 
catalogue had colours and magnitudes perturbed by the photometric errors and
typical weak lensing correction ({\it e.g.} if 20\% of the survey area was lensed
by 0.1 magnitudes, 20\% of the catalogue was randomly selected and the model magnitudes
were brightened.) Using this artificial catalogue, the galaxy candidates were reselected
using the original selection criteria ($I<26.3$, $R-I>1.3$ and $V>27.5$) to produce the 
predicted magnitude distribution for the ERGS survey for the assumed luminosity function 
with magnitude dependent completeness corrections (as described
in section \ref{sec:completeness}) in addition to the photometric errors, caused
by the ERGs image quality, and weak lensing corrections.

\begin{figure}
\begin{center}
\includegraphics[width=8cm]{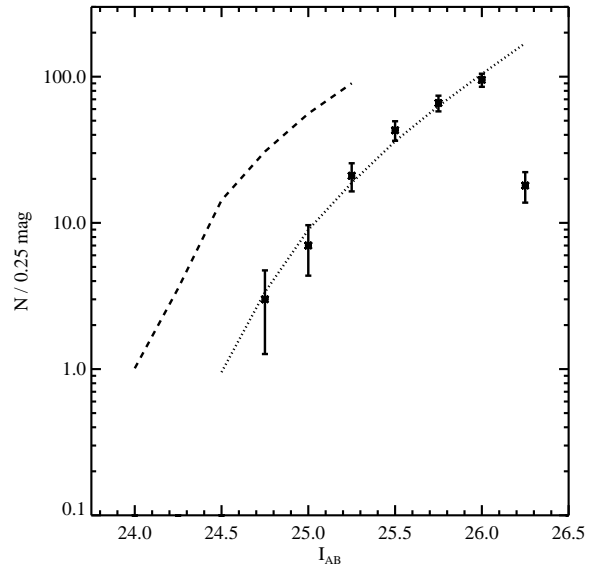}
\caption{The observed galaxy number counts across the 10 survey fields compared to the best fitting Schechter luminosity function (dotted line) with $\alpha=-1.6$ fixed. \citet{steidel99} $z\sim4$ luminosity function applied to $z=5$ assuming no evolution and given the imaging properties and selection of this study is shown by the dashed line. The faintest $I$-band magnitude bin was not included in the fitting procedure. The errors shown are the Poissonian errors of the number of objects in each bin.}
\label{fig:LF}
\end{center}
\end{figure}

Having repeated this procedure for each combination of luminosity function parameters,
 the predicted and observed number counts were compared using $\chi^2$ statistics across
the range of M$^*$ and $\alpha$, allowing the normalisation, $\phi^*$ to vary. The
resulting best fitting observed luminosity function is shown in figure \ref{fig:LF}. As this
method does not formally fit $\phi^*$, the correction for contamination is applied to the
normalisation after the fitting procedure.

\begin{figure}
\begin{center}
\includegraphics[width=8cm]{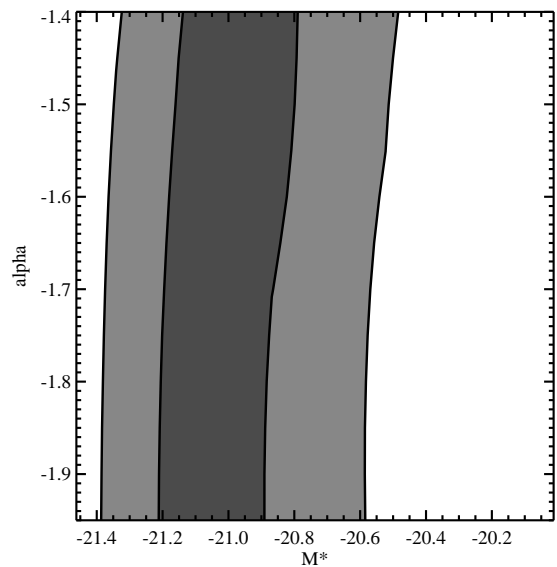}
\caption{$\chi^2$ space illustrating the fitting of Schechter functions to the observed galaxy number counts. The contours levels are 90\%, 95\% probabilities. With such a faint M* relative to the depth of the sample, the slope of the faint-end, $\alpha$, can not be constrained with this data.}
\label{fig:LFchi}
\end{center}
\end{figure}

As figures \ref{fig:LF} and \ref{fig:LFchi} show, our results for the
refined sample, are generally in 
agreement with those of \citet{bouwens07}. That result was derived
from extremely deep multi-band HST imaging of a comparatively small
area. The fidelity of the data means that it produces an intrinsically
cleaner sample with less contamination than typical ground-based
studies.  We determined a formal best-fitting M$^*=-20.96\pm{0.2}$ at
1500\AA, 0.3 magnitudes fainter than that found by
\citet{bouwens07} \citep[see also][for additional space and ground-based 
studies]{oesch07, beckwith06, yoshida06} with $\alpha=-1.4$.
 Given that the best-fitting M$^*$ falls into one of the
faintest bins used for model fitting, the data set in this paper can
not constrain the faint end slope (figure \ref{fig:LFchi}). When $\alpha$ was set to
-1.6 the $\chi^2$ best fitting characteristic magnitude was M$^*=-20.9\pm{0.2}$
and $\phi^{*}=1.1\pm 0.2 \times10^{-3}$ Mpc$^{-3}$ agreeing, within the errors, with
previous studies. As we are observing $z=5$ LBGs using an $I$-band filter at 7680\AA\,
the measured luminosity function is at 1250\AA\ compared to 1600\AA\ luminosity function
in \citet{bouwens07}. Such a difference in wavelength would result in different 
luminosity function parameters at lower redshift, due to the spectral slope of the source.
As high redshift LBGs have relatively flat continuum, particularly between 1250\AA\ and 1600\AA\
changes in the luminosity function parameters are expect to be small. 

Assuming the luminosity function fitted to the refined sample defines the shape of the relationship, 
the normalisation, as defined by $\phi^*$, can be adjusted to include the 
reliability of this sample ($\sim$80\%, see section \ref{sec:contamfrac}). 
The adjusted normalisation gives $\phi^{*}=0.9\pm 0.2 \times10^{-3}$ Mpc$^{-3}$
where the fraction of objects at high redshift is assumed to be constant with 
apparent magnitude. However, figure \ref{fig:ihist} suggests that contamination
is a more significant problem at brighter magnitudes with a high fraction of
lower redshift objects compared to high redshift candidates. A magnitude dependant 
contamination correction, with a larger correction in the brightest bins, 
would change the shape of the luminosity function leading
to a fainter M$^*$. Even with the extensive spectroscopic followup carried
out in the ERGS project, the number of confirmed high redshift
galaxies is not sufficient to study the reliability of the photometric
sample as a function of apparent magnitude.

For consistency with other work, we have used a \citet{schechter76}
luminosity function form for fitting. We note that the suitability of
an analytic form derived from the expected form of galaxy mass functions
is questionable for these intense starburst events.  The UV emission
arising from these LBGs is highly stochastic, tracing as it does short
lived intense star formation events. The characteristic timescale for
these is $\sim 30$Myr \citep[][]{verma07} in comparison to a survey
window of 250 Myr, {\it i.e.} the measured UV luminosity function is only
capturing a fraction of the overall LBG population. The UV luminosity
function is therefore likely to be weakly related to mass distribution
of star forming galaxies at these redshifts and therefore there is
little {\it a priori} reason to expect it to follow a Schechter
function. Indeed, it is plausible that the UV luminosity function is
shaped by the time-dependent decay of UV emission from these
short-lived starbursts, in effect the shape of the luminosity function
mirrors the UV emission (and therefore star formation) history of a
typical $z\sim 5$ LBG starburst. Although our data can be fit by a
Schechter function, it can be equally well fit by a simple power-law,
N(L)$\propto$L$^{-2.1}$. The main issue associated with determining
the detailed UV luminosity function from bright to faint
magnitudes at this redshift is the limited volume probed to the
faintest levels. This necessitates combining together the number
counts from at least two separate samples covering the bright and
faint ends of the function. As these come from different volumes, the
normalisation of the two ends can vary due to field-to-field
variation caused by cosmic variance, or by magnitude-dependent
contamination levels within the samples. Indeed, an intrinsic power
law extending from bright to faint levels could manifest an apparent
break if the bright and faint regimes were determined from separate
samples and the bright subsample was normalised high due to cosmic
variance or a higher level of magnitude-dependent contamination.

\subsection{Sources of Reionization}

\citet{lehnert03} claimed that the number of $z>5$ sources, selected in
a similar manner to this work, was less than that expected from the
$z=4$ luminosity function of Lyman Break galaxies (about 1/3 the
number of sources for $I<26.3$), assuming no evolution between the epochs.
 Although their data was taken with the same instrument
as used in this work, their conclusion was drawn from only 10\% of the sky
area of this survey. Nevertheless, our average number
counts over 10 fields also implies a similar lack of bright sources compared
to that expected if there was no evolution in the luminosity function
between $z=5$ and $z=4$.

Figure \ref{fig:LF} compares the observed galaxy number counts
(crosses) with that predicted using \citet{steidel99} $z=4$ luminosity
function down to $M^*$. It can be seen that the observed number counts are a factor of $\sim$3 less than
would be expected if there is no evolution in the luminosity function
between $z\sim5$ and $z\sim4$, a deficit which increases with
luminosity. Although the epoch of reionization has ended by $z\sim5$,
the Universe still requires a sufficient density of UV photons to
counteract the process of recombination and keep the hydrogen
ionised. Even without the removal of contaminants, there is a lack of
galaxies providing ionising photons at these magnitudes. 

The equation for the density of star-formation required for reionization
from \citet{madau99}, modified to the cosmology assumed in this study, was used
to estimate the star-formation density required to maintain the ionisation 
state of hydrogen at a redshift of 5.1, the mean redshift of the
spectroscopically confirmed galaxies within the photometric sample. The concentration factor of hydrogen
was assumed to be 30 and the photon escape fraction used was $f_{esc}=0.14$
\citep{shapley06}, however these values are not well constrained and can vary largely.
The total luminosity of the sample was found by 
converting all apparent magnitudes
assuming the mean redshift of the confirmed high
redshift galaxies within the sample for all objects. This was then converted 
to a star-formation rate using the relation given in \citet{madau98}.

It was found that the star-formation density of the refined sample is
a factor of $\sim$4 lower than that necessary to maintain
the ionisation of hydrogen at redshift $\sim$5. Even when the whole photometric sample is
used, without the removal of likely contaminants, the star-formation density is still nearly
3 times lower than required. These results imply that the
majority of ionising photons originated from fainter sources below the
observed flux limit, \citep[$I>26.3$;][]{lehnert03} even after the
Universe has fully reionized. Another potential source of ionising
photons are AGN. However, \citet{bremer04} and \citet{douglas07} have
shown that the density of AGN at these redshifts is very low and as
such can not be a large contributor to the overall photon budget of
the Universe at these redshifts.

\section{Conclusions}
\label{sec:conclusions}

Through a photometric colour selection we have produced a sample of
253 high redshift Lyman Break Galaxy candidates at $z\sim5$ drawn from 10
widely-separated fields with $V,R,I,z,J,K_{s}$ photometry covering 275
arcmin$^2$.  Using simulations and a comparatively large programme of
follow-up spectroscopy we have been able to determine the completeness
and reliability of the sample. The sample of
253 sources is approximately 80 per cent reliable ({\it i.e.} has a
$\sim20$ per cent contamination rate). Even with multiple photometric bands used to
select objects, the relatively extreme colours of dropout galaxies
make them vulnerable to systematic effects in photometry of individual
objects. Consequently, statistical results derived from large samples
of photometrically-selected objects may well be prone to bias or error
because of contamination within the samples. As has been pointed out
elsewhere \citep[e.g.][]{stanway08b}, contamination rates of only 15
per cent can significantly skew some statistical
results. 

A large fraction of these galaxies have multiple components on scales
probed by HST imaging. It is currently unclear whether the components are
separate galaxies, or UV luminous regions embedded in a larger, darker
underlying galaxy but deeper HUDF studies suggest we are observing 
multiple galaxy systems. Previous work has shown that LBGs appear to
decrease in linear size with increasing redshift \citep{bouwens04}. This applies to
individual components of multiple systems, but clearly not necessarily
the systems themselves.

Contamination is a potentially serious issue in constraining the
$z\sim 5$ luminosity function. Having taken into account our detailed
reliability analysis, we confirm a decline of a factor of $\sim 3$
in the bright end of the luminosity function (to $M^*$) from $z=4$ to $z=5$. Assuming a Schechter fit to the UV
luminosity function at $z\sim5$, we obtain values for $M^*_{UV}=-20.9\pm 0.2$ and
$\phi^*=0.9\pm 0.2 \times 10^{-3}$ Mpc$^{-3}$ (with a magnitude independent correction 
for completeness), in
agreement with \cite{bouwens07} and others. However, we question the basis for the use of a Schechter
function. This is a proxy for a mass function, whereas the UV
luminosity of LBGs arises from highly stochastic, short-lived
starburst phases and should not necessarily correlate with the mass of
the underlying system.

\section*{Acknowledgements}
We thank the referee for their helpful comments. 
LSD acknowledges support from P2I. ERS acknowledges support from STFC.
Based on observations made
with ESO Telescopes at the La Silla and Paranal Observatory under
programme IDs 166.A-0162 and 175.A-0706. Also based on observations made
with the NASA/ESA Hubble Space Telescope, obtained at the Space
Telescope Science Institute, which is operated by the Association of
Universities for Research in Astronomy, Inc., under NASA contract NAS
5-26555. These observations are associated with program 9476. We thank
the members of the EDisCS collaboration for creating an imaging
dataset with a usefulness far beyond their original intent.

\label{lastpage}

\end{document}